\documentstyle[epsfig]{aipproc}
\font\sc=cmbx10 
\font\smcap=cmcsc10 at 12pt 
\font\bigbf=cmbx12 at 14.4pt
\font\hugebf=cmbx12 at 17.3pt
\def\kps{km s$^{-1}$}
\def\B{$B$}
\def\R{$R$}

\def\hi{\smcap H$\,$i}

\def\et{{\it et al.}}
\def\etal{{\it et al.}}
\def\micron{$\mu$m}

\def\spose#1{\hbox to 0pt{#1\hss}}
\def\simlt{$\mathrel{\spose{\lower 3pt\hbox{$\mathchar"218$}}
\raise 2.0pt\hbox{$\mathchar"13C$}}$}
\def\simgt{$\mathrel{\spose{\lower 3pt\hbox{$\mathchar"218$}}
\raise 2.0pt\hbox{$\mathchar"13E$}}$}

\begin{document}

\hfil{}
\vfill
\centerline{\hugebf Mergers, Interactions,}
\centerline{\hugebf and The Fueling of Starbursts}
\bigskip
\centerline{\bf J. E. Hibbard\footnote{Hubble Fellow}}
\smallskip
\centerline{\it Institute for Astronomy, University of Hawaii}
\smallskip
\centerline{\it hibbard@uhifa.ifa.hawaii.edu}


\vfill
\centerline{To appear in {\it ``Star Formation, Near and Far"},}
\centerline{The 7$^{th}$ Annual Astrophysics Conference in Maryland,}
\centerline{S.S.~Holt \& L.G.~Mundy, editors.}

\bigskip\bigskip
\centerline{Electronic version of this paper is available at:}
\centerline {\it http://www.ifa.hawaii.edu/$\sim$hibbard/MdConf/}

\vfill
\noindent{\bf U{\sc NIVERSITY OF} H{\sc AWAII}}\hfill\break
{\bf INSTITUTE FOR ASTRONOMY}\hfill\break
{\bf 2680 Woodlawn Drive}\hfill\break
{\bf Honolulu, Hawaii 96822 USA}
\vfill\eject
\hfil{}\vfill\eject

\title{Mergers, Interactions,\\ and The Fueling of Starbursts}

\author{John E. Hibbard}
\address{Institute for Astronomy\\
2680 Woodlawn Drive\\
Honolulu, Hawai'i  96822}

\maketitle

\vspace{-10pt}
\begin{abstract} The most active starbursts are found in galaxies with
the highest IR luminosities, with peak star formation rates and
efficiencies that are over an order of magnitude higher than in normal
disk systems.  These systems are almost exclusively on-going mergers. 
In this review I explore the conditions needed for interactions to
experience such a phase by comparing two systems at similar stages of
merging but quite different IR luminosities: NGC 4038/9 and Arp 299. 
These observations show that the most intense starbursts occur at the
sites with the highest gas densities, which is a general result for IR
luminous mergers.  Observations and theory both suggest that the
strength of the merger induced starburst depends on the internal
structure of the progenitors, the amount and distribution of the gas,
and the violence of the interaction.  In particular, interactions
involving progenitors with dense bulges, gas-rich disks, and/or a
retrograde spin are expected to preferentially lead to large amounts of
gaseous dissipation, although the interplay between these parameters is
unknown.  A major outstanding question is how the effects of feedback
alter these conclusions.  \end{abstract}

\section*{Introduction}

While galaxies showing both peculiar morphologies and signs of global
youth have been known for decades
(e.g.~\cite{Zwicky50,Burbidge63,flashing}), it was not until a classic
study on the $UBV$ colors of peculiar galaxies by Larson \& Tinsley in
1978 \cite{LT} that the study of starbursts in interacting galaxies
began in earnest.  In that seminal work the authors showed that the
disturbed systems from the Arp {\it Atlas of Peculiar Galaxies}
\cite{ArpAtlas} had a larger spread in colors and significantly bluer
colors than a comparison sample of normal Hubble types.  The Toomres
\cite{TT} had recently demonstrated quite convincingly that
gravitational interactions provide a natural explanation for many of the
types of peculiarities exhibited by the systems in Arps atlas, and
Larson \& Tinsley posited that such interactions induce small bursts
($\sim$ 1-5\% by mass) of star formation within the host(s).  They were
able to explain the color distribution of the Arp systems with burst
models of varying strength and age superimposed upon an underlying host
with normal colors. 

\begin{figure}[t!] 
\centerline{\epsfig{file=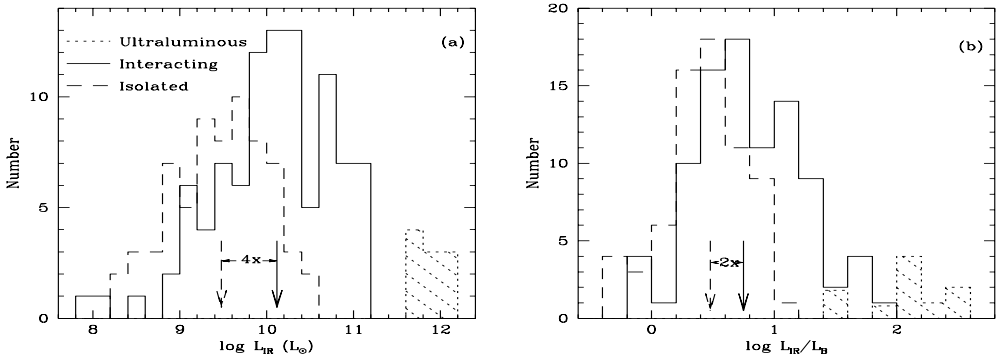,height=2in,width=5.5in}}
\vspace{10pt}
\caption[]{Histograms of $L_{IR}$ and $L_{IR}/L_B$ for isolated, interacting,
and infrared bright systems. The isolated and interacting samples were
chosen optically, independent of their infrared properties. Arrows indicate
their means. These plots show that while the brightest IR emitting systems 
are interacting, many interacting systems are not IR bright. From 
Bushouse, Lamb \& Werner 1988 \cite{BLW}.
}\label{fig1}
\vspace{-10pt}
\end{figure}

Much subsequent work has supported this suggestion \cite{Kennicutt90}. 
The consensus is that interacting galaxies as a class have mean levels
of star formation that are factors of 2--5 higher than normal spirals
for optically selected samples, or 2--20 times higher for samples
selected on the basis of IR luminosity.  This point is illustrated in
Figure \ref{fig1} \cite{BLW}, which compares the IR luminosities of an
optically selected samples of isolated and interacting galaxies.  Also
shown are the values for a sample of ultraluminous infrared
(ULIR\footnote{$L_{IR}>3\times10^{11} L_{\odot}$ in this case, although
other definitions are used}) galaxies which are known to be on-going
mergers\cite{Sanders88}.  Similar results have been derived using other
measures of star formation, finding that interactions serve primarily to
concentrate moderately enhanced star formation into the central regions
of galaxies rather than to globally raise the star formation rate (e.g. 
\cite{Keel85,Bushouse87,Kennicutt87,Condon91}). 

Figure \ref{fig1} illustrates several other points.  The first is that
none of the isolated systems have $L_{IR}>3\times10^{10} L_{\odot}~{\rm
or}~L_{IR}/L_B>15$.  At these levels almost all galaxies are interacting
or merging \cite{Sanders88,Lonsdale84,JW85}.  In fact, IR luminosity
appears to be the most efficient way to select interacting systems:
while the overall peculiar fraction of optically selected samples is
around 9\% \cite{ArpMadore}, the fraction of morphologically peculiar
galaxies approaches 90\% or higher at IR luminosities above $5\times
10^{11} L_{\odot}$ (see \cite{SMaraa} and references therein).  The
second is that many systems in the interacting sample do {\it not}
exhibit enhanced levels of star formation.  So while the most luminous
IR sources are almost invariably mergers, not all mergers are luminous
IR sources. 

IR luminosity is believed to be directly related to the number of hot
stars present, as the dust absorbs the UV photons from these stars and
reradiates them in the IR \cite{Lonsdale84,JW85}.  It can therefore be
used as a measure of the massive star formation rate (MSFR; $M>5
M_\odot$) \cite{CondonARAA}.  The IR luminous systems are 
both gas-rich and dusty\cite{Sanders88,Sco89,Sco91}, and most of their
bolometric luminosity emerges in the far IR \cite{Sanders88,SMaraa}. 
While there has been a long-standing debate over whether the most
luminous systems are instead powered by an obscured AGN (see
e.g.~\cite{Condon91,Lonsdale95}), this ambiguity is mostly at the
highest IR luminosities, and the majority are believed to be
predominantly powered by starbursts (see reviews by
\cite{Kennicutt90,Heckman90}).  The popular picture which has emerged is
that two gas rich systems undergo a close interaction, which leads to
orbital decay and eventual merging.  The gas is compressed, dissipates
and moves inward, stimulating a circumnuclear starburst.  Dust, which is
coupled to the gas, absorbs much of the UV radiation and re-radiates it
in the IR.  This high level of star formation quickly subsides as the
burst consumes the available gas. 

But do all mergers experience a strong burst of star formation? If not,
what are the deciding factors? These are the issues I will discuss in
this review. 

\begin{figure}[t!] 
\centerline{\epsfig{file=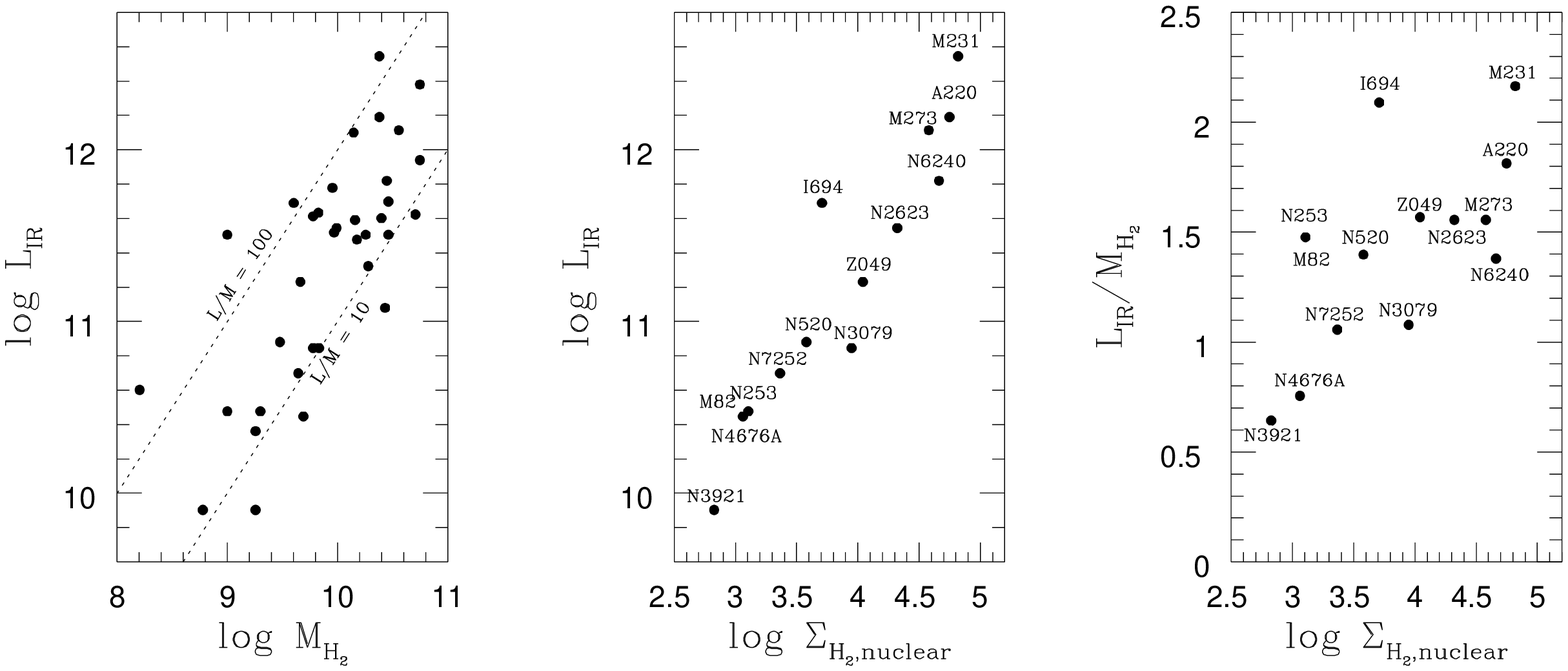,height=2.5in,width=5.5in}}
\vspace{10pt}
\caption{{\bf (a)} total IR luminosity vs.~derived $H_2$ mass for all
objects observed at OVRO.  Normal spirals fall below the L/M=10 line. 
{\bf (b)} $L_{IR}$ vs.~the molecular gas column density averaged over a
1 kpc diameter region ($\Sigma_{H2,nuclear}$) for the 12 objects
observed with sufficient resolution.  {\bf (c)} SFE
vs.~$\Sigma_{H2,nuclear}$.  From Yun \& Hibbard, in preparation.}
\label{fig2}
\vspace{-10pt}
\end{figure}

\bigskip
\centerline{\bigbf Quantifying the Star Formation Activity}
\medskip

Although star formation activity has historically been defined by some
absolute quantity such as IR luminosity, it would be preferable to use a
relative measure which compares the total SFR to the the size of the
system.  In this way we can address whether star formation proceeds in
an inherently different manner at these luminosities, or if it is 
a scaled up version of processes taking place at lower luminosities. 
Blue luminosity is frequently used as a substitute for the stellar mass
(e.g.~Fig.~1b), although this interpretation is wrought with
uncertainties due to the presence of young stars and large amounts of
dust.  $L_{IR}/L_K$ is a more useful measure, although $L_K$ is still
affected by the presence of red AGB stars. 

A popular normalization introduced by millimeter astronomers is the
ratio of the IR luminosity to the molecular gas mass ($L_{IR}/M_{H2}$),
where $M_{H2}$ is estimated from measurements of the CO line.  This
ratio is termed the ``star formation efficiency" (SFE)
\cite{Sco91,Young86,SolSage} as it represents in some global
sense the number of massive stars formed per giant molecular cloud. 
Since $L_{IR}\sim$MSFR \cite{CondonARAA}, the SFE is inversely related
to the gas depletion time, and is thus equivalent to the more classical
definition of a starburst ({\it i.e}~$\Delta t_{burst} << H_o^{-1}$). 


Many studies have evaluated the SFE in merging galaxies concluding that
ULIR mergers are forming stars up to an order of magnitude more
efficiently than normal spirals (see Figure \ref{fig2} and
\cite{SMaraa}).  While there is some question as to whether the Galactic
conversion factor between CO and H$_2$ applies in merging galaxies
\cite{Sco89,Maloney,DSR}, the expected variation leads to an
overestimate of the molecular gas and therefore underestimates the
actual SFEs.  It therefore seems hard to escape the conclusion that
massive stars are being formed at a higher rate for a given amount of
cold gas than in quiescent systems, although the exact level of the
enhancement is uncertain. 

\bigskip
\centerline{\bigbf The Tails of Two Mergers}
\medskip

To understand how this efficient mode of star formation is triggered, it
is instructive to compare two nearby mergers at apparently similar
stages of merging but at two quite different levels of star formation
activity.  The mergers I will examine are ``The Antennae" (NGC 4038/9,
Arp 244; $V_o$=1630 \kps) and Arp 299 (NGC 3690/IC 694; $V_o$=3080
\kps).  Both the tidal structure and inner disks of these systems are
shown in Figure \ref{fig3} along with distribution of the cold gas
components. 

\begin{figure}[tp] 
\vspace{6.66 truein}
\caption[]{Two on-going mergers. (a)\&(c) NGC 4038/9 and (b)\&(d) Arp 299.
In the upper two panels, a deep \R-band image of the entire system is 
shown with VLA {\hi} contours overlaid, while in the lower two panels 
a \B-band image of the inner regions is shown with OVRO $^{12}$CO(1-0) 
contours overlain. The optical data were obtained by the author; the
VLA data are from Hibbard \& Yun 1996 \cite{HY96} (Arp 299) 
and Hibbard \& van der Hulst in preparation (NGC 4038/9); the OVRO data 
are from Aalto \et\ 1997 \cite{Aalto}
(Arp 299) and Stanford \et\ 1990 \cite{n4038co} (NGC 4038/9). These 
systems appear to be at similar stages of merging, with their disks in 
contact and their nuclei still separate. 
}\label{fig3}
\end{figure}

The evidence that these systems are at a similar stage of merging is the
following: both have a long tidal tails (130 kpc for NGC 4038/9 assuming
a distance of 25 Mpc; 180 kpc for Arp 299 assuming a distance of 48
Mpc), suggesting several hundred Myr since first orbital periapse
(\simgt 500 Myr for NGC 4038/9; \simgt 700 Myr for Arp 299); the disks
of their progenitors are highly distorted and in physical contact, yet
still distinct; and their nuclei are still well separated (8 kpc and 4
kpc, respectively).  Both systems have similar CO distributions
(Fig.~3c-d), with concentrations of cold molecular gas near both nuclei
and a significant concentration at the region of disk overlap.  Despite
these similarities, Arp 299 is almost an order of magnitude more
luminous in the infrared than NGC 4038/9, indicating a much higher MSFR
(see Table~1). 

\begin{table}[t!]
\caption{Properties of Star Forming Regions in Arp 299 \& NGC 4038/9.}
\label{table1}
\begin{tabular}{lcccccc}
  & $L_{IR}$\tablenote{$L_{IR}$ is split between components 
  using 15\micron\ ISOCAM measurements for NGC 4038/9 \cite{ISO}, and 
  32\micron\ measurement for Arp 299 \cite{a299}.} 
  & $M_{H2}$\tablenote{From OVRO CO observations of NGC 4038/9 \cite{n4038co}
  and Arp 299 \cite{a299co}.}
  & $L_{IR}/L_K$ & MSFR & $\Sigma_{H2}$ & SFE \\
  & ($\times10^{10} L_{\odot})$ & ($\times10^{9} M_{\odot}$) 
  & & ($M_{\odot} {\rm yr}^{-1})$ & ($M_{\odot} {\rm pc}^{-2}$) 
  & ($L_{\odot} M_{\odot}^{-1}$) \\
\tableline
IC694   & 49.8 & 4.0 & --- & 31 & 27,000 & 124  \\
N3690   & 24.5 & 1.0 & --- & 15 &  4,900 & 245  \\
overlap &  4.7 & 2.0 & --- &  3 &  3,400 &  24 \\
\tableline
\tableline
Arp 299 & 79.0 & 7.0 &  31 & 49 & 27,000 & 113 \\
\\
N4038   &  2.0 & 0.8 & --- &  1 &  1,200 &  25  \\
N4039   & 0.04 & 0.2 & --- &  0 &    540 &   2  \\
overlap &  7.8 & 1.2 & --- &  5 &  1,350 &  65  \\
\tableline
\tableline
N4038/9 & ~9.8 & 2.0 &  13 &  6 &  1,350 &  49 \\
\end{tabular}
\end{table}

In Table 1 we use NIR flux measurements \cite{ISO,a299} to divide the
IR luminosity among the various components.  This comparisons shows
that a major difference between the systems is in the amount of
nuclear star forming activity: in NGC 4038/9, the overlap region is the
most active star forming region, both in terms of the MSFR and the SFE. 
In Arp 299, on the other hand, the overlap region has similar properties
as that in NGC 4038/9, but both nuclei outshine this region by large
amounts.  We note that observations argue against an energetically
dominant AGN in any of the nuclei \cite{Condon91,ISO,a299,Ridgway}
(but see \cite{TopHeavy}). 

Similar differences are seen in the column densities of molecular gas:
there is five times as much gas in the central kpc of the nuclei of Arp
299 as in NGC 4038/9.  As a result, the peak column densities are over
an order of magnitude higher in the nuclei of Arp 299 as in the nuclei
of NGC 4038/9 or the overlap regions.  The tight correlation between
$\Sigma_{H2}$ and $L_{IR}$ shown in Fig.~\ref{fig2} shows this to be a
general result for mergers over a broad range of IR luminosity
\cite{Sco91}.  While it is still possible that $M_{H2}$ has been
overestimated \cite{Maloney,DSR} a similarly tight correlation exists
between $L_{IR}$ and $L_{HCN}$, as well as between SFE and
$L_{HCN}/L_{CO}$ \cite{SDR}, showing that the fraction of gas at very
high densities ($n>10^5 {\rm cm}^{-3}$) is greatly increased in ULIR
systems.  As these fractions are much higher than those found in normal
spirals, this indicates that MSFR is intricately linked to the amount of
gaseous dissipation \cite{Sco89,DSR,KS}. 

It therefore seems that in order to understand how such efficient
periods of star formation are induced, we need to understand how so much
gas attains such high gas densities.  Most phenomenological scenarios of
star formation predict that the MSFR and the SFE are proportional to the
gas density to some power (e.g.~\cite{Silk,Elmegreen}), although they do
not address in detail when and where such densities are attained.  For
this, we turn now to results from numerical simulations. 

\bigskip
\centerline{\bigbf Lessons from Simulations}
\medskip

Ever since a dissipational component was first included in numerical
simulation of interactions it has been clear that large amounts of gas
can be efficiently driven into the central regions \cite{NW,NI,BH91}. 
In this section I examine individual models for clues as to what
parameters play the largest role in driving gas to high densities.  This
is not intended to be a review of all the simulations (for that see
\cite{BHaraa}), but rather an assessment of the results which might have
the most relevance for understanding the observations discussed above. 

\begin{figure}[t!] 
\vspace{4.51 truein}
\caption[]{Time history of two interaction induced starbursts.  In the
bottom sequence, the lack of a dense bulge allows strong bar-induced 
inflows to develop early in the encounter. The relatively mild 
starburst which ensues consumes much of the available gas supply, 
leading to a less extreme burst at the final merger.  Adapted from 
Mihos \& Hernquist \cite{MH94}.}
\vspace{-12pt}
\label{fig4}
\end{figure}

\paragraph*{Progenitors:} A very interesting result to emerge from the
numerical studies is that the progenitor structure ought to play a
dominant role in the star formation history (SFH) \cite{MH94,MH96}. 
This is illustrated in Figure \ref{fig4}.  In the upper sequence the
presence of a bulge stabilizes the disks against bar formation, lowering
the pre-merger star formation levels, but leaving more gas in the
progenitors to fuel a strong starburst when they finally merge.  Because
of the peaked appearance, it is tempting to associated this SFH with the
ULIR systems, but it is important to realize that the vertical axis is a
{\it relative} SFR, and the absolute SFR between the two curves could
differ due to other factors (see below).  Indeed, the fact that Arp 299
is so luminous while the nuclei are still well separated suggests that
it follows the lower SFH in Fig.~\ref{fig4}.  This interpretation is
supported by the {\hi} observations, which suggests two late type
progenitors \cite{HY97}.  NGC 4038/9 on the other hand appears to have
one late type (NGC 4038) and one earlier type (NGC 4039) progenitor
\cite{vdH79}, and may have a SFH more like the upper curve in
Fig.~\ref{fig4}, with its most active star forming period still to come. 
These considerations suggest that while a dense bulge may push certain
systems into the ULIR regime, it need not be a strict requirement.  They
also point to the need for better constraints on the past and future
star forming histories to facilitate comparisons with models. 


\paragraph*{Gas content:} There is over three times as much molecular
gas in Arp 299 than NGC 4038/9, although the relative gas contents are
similar ($M_{H2}/L_K$=0.5, 0.4).  Could this be responsible for the
different levels of activity? Clearly the level of star formation must
be closely connected to the amount of fuel available, but plots of SFE
or $L_{IR}~vs.~M_{H2}$ show notoriously large amounts of scatter
(e.g.~Fig.~2a).  So while the gas content may help shift the SFR upward
for some systems (see Gao \et\ these proceedings), high gas mass alone
is not a sufficient condition for fueling an ULIR phase.  Olson \& Kwan
\cite{OK} conducted one of the few simulations to explore the effects of
different gas masses and distributions.  Using a code in which the SFR
is proportional to the cloud collision rate, they find that doubling the
gas content in one of the disks in a disk-disk merger doubles the SFR,
therefore keeping the SFE (=SFR/$M_{gas}$) constant, while splitting the
same amount of gas between two galaxies leads to a 30\% higher SFR and
SFE than putting it all into one of the system.  This suggests that how
the gas is distributed and how it collides is as important as how much
gas is present.  These studies should be continued with a wider range of
encounter parameters. 

\paragraph*{Spin geometries:} The two systems under consideration here
differ in their spin geometries, with Arp 299 undergoing a
prograde-retrograde encounter \cite{HY97,a299spin} and NGC 4038/9
undergoing a prograde-prograde encounter \cite{vdH79}.  Since retrograde
encounters fail to raise strong tails \cite{BH96}, there will be more
gas left in the inner regions at the late stages of merging where it
will be violently perturbed as the encounter progresses.  The
simulations of Barnes \& Hernquist \cite{BH96} show that the fraction of
gas at high densities depends on encounter geometry, with retrograde
encounters leading to higher quantities of dense gas than prograde
encounters.  In Arp 299, the highest gas column densities are found in
the nucleus of the disk experiencing the retrograde encounter (IC 694). 
Mihos \& Hernquist \cite{MH94,MH96} also evaluated the effects of spin
geometry, and while they showed that it had a much less dramatic effect
on the SFRs than progenitor structure, geometry still made a difference
of about a factor of two in the relative SFRs.  If there is either a
threshold to the onset of very efficient star formation activity, or if
SFE is a function of the fraction of the gas at very high densities, it
is possible that spin geometry plays a role beyond that attributed to it
in these studies. 

The reality is that all of the factors are probably playing a part,
although in what combination and order of importance is not clear. 
Continued parameter studies are needed in concert with detailed
observations of individual systems in order to discriminate between the
different processes.  It is also important to compare the actual
distributions and dynamics of the star forming regions and dense gas
with the predictions of the simulations in order to discriminate between
different numerical formalisms for star formation and gas dynamics. 
Only by doing this will we know how far to trust the models or how to
better conduct our observations. 

\bigskip
\centerline{\bigbf Other Aspects of Interaction Induced Starbursts}
\medskip

There are many other interesting aspects of interaction induced starbursts,
and in the remaining space I simply mention a few.

\paragraph*{Star Formation Knots:} When viewed with sufficient
resolution the star-forming regions in Arp 299 and NGC 4038/9 are found
to break up into many distinct knots along with a diffuse component. 
Figure \ref{fig5} shows an $HST$~$FOC$ image of NGC 3690
\cite{Meurer,Vacca95}.  These knots have typical diameters of less than
10pc and may evolve into globular clusters \cite{Meurer,n4038gc}. 
However these knots are not unique to mergers
\cite{Meurer,CV,Holtzman,Maoz}, and it is not clear if this mode of star
formation is increased in such interactions and if it is related to the
enhanced SFE.  It may just be an important mode of star formation in
general.  A detailed comparison of the luminosity functions of such
knots from different types of environments would help resolve this
question. 

\begin{figure}[t!] 
\vspace{2.26 truein}
\caption[]{$HST~FOC~2200 \AA$ observation of the B-C-C$^\prime$ complex
in NGC 3690 (Arp 299 West) from Vacca 1995 \cite{Vacca95}. 
These observations show that many of the massive stars in this system 
are confined to very bright, compact knots. 
}
\label{fig5}
\vspace{-10pt}
\end{figure}

\paragraph*{The Return of Tidal Debris:} Due to the strong tides
experienced during merging encounters, appreciable amounts of stars and
gas are lifted high above the merging systems into tidal features. 
These features frequently exhibit significant substructure, the largest
having observational properties typical of dwarf galaxies
\cite{S78,TDs,me94,me96}.  Much of this material remains bound to the
remnant on long-period orbits, and will take several Gyr to fall-back
\cite{HS92,me95}.  Tidal clumps that are far out along the tails may be
able to avoid tidal stripping when they fall back towards the remnant
and should become long-lived dwarf companions\cite{me95}, while the more
tightly bound material will fall back into the remnant.  The stellar
component of these tails will wrap coherently in the central potential,
giving rise to fine structure features \cite{HS92} while the gas may
feed a prolonged period of low-level star formation.  The largest clumps
have as much as a few 10$^8 M_{\odot}$ of {\hi} \cite{me94,me96}, and
their return may give rise to smaller bursts.  The overall star
formation history in merger events should be similar to that illustrated
by Worthy (this volume). 

Because of the timescales and amounts of gas involved, we therefore do
not expect two merging spirals to turn quickly into an elliptical, but
rather for there to be a series of transitions, e.g.~to an S0pec, to an
S0, to a dust lane elliptical, etc.  \cite{me96}.  By the time the more
obvious signs of its merger origin have faded and the remnant has
evolved into a {\it bona fide} elliptical, the stars formed during the
merger induced starburst will have aged 2-5 Gyr, leaving very little
indication of a merger origin in the remnants broad band colors.  This
picture is quite similar to the one emerging from studies of cluster
populations at high redshift (see Dressler, these proceedings). 

\bigskip
\centerline{\bigbf Outstanding Questions}
\medskip

The main outstanding question for interaction induced star formation is
the same as for any field of star formation: is the IMF universal, and
if not how is it different in these violent environments? Are there
upper and lower mass limits? Is the IMF top heavy? All of these effects
have been claimed \cite{TopHeavy,HiCut,LowCut}, but the observations
allow for a wide range of parameters.  A major problem with these
determinations is that the observered bursts are both temporally and
spatially variable, and as a result there will be a mix of burst
populations of varying strength and ages spread throughout the merger. 
An IMF measured globally will reflect the luminosity averaged IMF over
the region observed, which will be different in different wavebands. 
Careful UV and/or NIR spectroscopy of individual knots in conjunction
with dynamical modeling should help separate these effects.  Such
observations should also help constrain the past star formation history
in the systems, allowing one to map the age distribution of the star
forming episodes. 

Another major question is how does the energy injected back into the
surrounding gas via the winds from massive stars and SNe change the gas
dynamics and burst properties? This process, referred to as ``feedback",
is seen in its most extreme form in mergers, as evidenced by the
galactic scale superwinds emerging from many ULIR systems (see Heckman,
these proceedings).  It may simply be an interesting side-effect of the
circumnuclear star bursts, or it may dramatically affect the physics of
star formation, for example by raising the lower-mass cutoff of the IMF
or changing its high-end slope \cite{ZepfSilk}, or by regulating the
star formation rate at some critical value \cite{Lehnert96}. 

There is much hope for further progress to be made in these areas in the
future.  It is becoming feasible to model a multi-phase ISM, which
should help assess the importance of feedback.  Further numerical trade
studies should help decide how the different parameters interact and
predict relationships between gaseous and star forming regions.  Careful
observations can test these predictions, and in this way we can
discriminate between the different numerical formalisms.  Not only will
this provide a deeper understanding of interaction induced star
formation, but ultimately it should be possible to decide which of the
two histories depicted in Fig.~\ref{fig4} a merger follows, and which of
these lead to an ULIR phase.  Since the ULIR systems may be the closest
analogs to the star forming galaxies seen at high redshift (see Madau,
these proceedings), this understanding will provide valuable insight
into the major processes at work during the epoch of galaxy formation. 

I wish to thank J.~van Gorkom and W.~Vacca for comments on this
manuscript; to M.~Yun, C.~Mihos, S.~Aalto and W.~Vacca for permission to
reproduce their figures; and to the organizing comittee for setting up
such an interesting meeting.  This work is supported by Grant
HF--1059.01--94A from STScI, which is operated by AURA, Inc., under NASA
contract NAS5--26555.

\end{document}